# An Efficient Convolutional Neural Network for Coronary Heart Disease Prediction

Aniruddha Dutta[1, 2], Tamal Batabyal[3,6], Meheli Basu[4], Scott T. Acton[3, 5]

[1]Department of Pathology & Molecular Medicine, Queen's University, Kingston, ON K7L 3N6, Canada
[2]Haas School of Business, University of California, Berkeley, CA 94720, USA
[3]Department of Electrical & Computer Engineering, University of Virginia, VA 22904, USA
[4]Katz Graduate School of Business, University of Pittsburgh, PA 15260, USA
[5]Department of Biomedical Engineering, University of Virginia, VA 22904, USA
[6]Department of Neurology, School of Medicine, University of Virginia, VA 22904, USA

**Abstract:**

This study proposes an efficient neural network with convolutional layers to classify significantly class-imbalanced clinical data. The data is curated from the National Health and Nutritional Examination Survey (NHANES) with the goal of predicting the occurrence of Coronary Heart Disease (CHD). While the majority of the existing machine learning models that have been used on this class of data are vulnerable to class imbalance even after the adjustment of class-specific weights, our simple two-layer CNN exhibits resilience to the imbalance with fair harmony in class-specific performance. Given a highly imbalanced dataset, it is often challenging to simultaneously achieve a high class 1 (true CHD prediction rate) accuracy along with a high class 0 accuracy, as the test data size increases. We adopt a two-step approach: first, we employ least absolute shrinkage and selection operator (LASSO) based feature weight assessment followed by majority-voting based identification of important features. Next, the important features are homogenized by using a fully connected layer, a crucial step before passing the output of the layer to successive convolutional stages. We also propose a training routine per epoch, akin to a simulated annealing process, to boost the classification accuracy.

Despite a high class imbalance in the NHANES dataset, the investigation confirms that our proposed CNN architecture has the classification power of 77% to correctly classify the presence of CHD and 81.8% to accurately classify the absence of CHD cases on a testing data, which is 85.70% of the total dataset. This result signifies that the proposed architecture can be generalized to other studies in healthcare with a similar order of features and imbalances. While the recall values obtained from other machine learning methods, such as SVM and random forest, are comparable to that of our proposed CNN model, our model predicts the negative (Non-CHD) cases with higher accuracy. Our model architecture exhibits a way forward to develop better investigative tools, improved medical treatment and lower diagnostic costs by incorporating a smart diagnostic system in the healthcare system. The balanced accuracy of our model (79.5%) is also better than individual accuracies of SVM or random forest classifiers. The CNN classifier results in high specificity and test accuracy along with high values of recall and area under the curve (AUC).

*Keywords:* Coronary heart disease, Machine learning, LASSO regression, Convolutional neural network, Artificial Intelligence, NHANES

**1. Introduction**

Heart disease is a leading cause of death today, with coronary heart disease (CHD) being the most common form of cardiovascular disease that accounts for approximately 13% of deaths in the US (Benjamin, 2019). Timely diagnosis of heart disease is crucial in reducing health risk and preventing cardiac arrests. An American Heart Association study projects an almost 100% increase in CHD cases by 2030 (Benjamin, 2019). Major risk factors such as smoking, hypertension, hyper cholesterol and diabetes have been studied in connection to CHD (Kannel, et al., 1971; Stamler, et al., 1993; Vasan, et al., 2001; Kannel, 1996; Burke et al., 1997; Celermajer et al., 1993; Chobanian et al., 2003; Haskell et al., 1994; Zeiher et al., 1993; Ahmed et al., 2017). Ahmed. et al. (2017) show that Body Mass Index (BMI) and systolic blood pressure are the two most critical factors affecting hypertension. Fava. et al. (2013) conclude significant association between age, sex, BMI and heart rate with hypertension. Studies in general population indicate that high level of creatinine in blood can increase the risk of CHD (Irie et al., 2006; Wannamethee, Shaper & Perry 1997). Additionally, blood cholesterol and glycohaemoglobin levels are found to be persistently and significantly high in patients with CHD (Burchfiel, 1997; Meigs et al., 1997). Several researchers have used statistical and machine learning models on echocardiography images (Nakanishi et. al., 2018; Madani et. al., 2018) and electrocardiography signals (Jin, 2009; Shen et. al., 2016) to predict clinically significant parameters related to CHD in patients, such as heart rate and axis deviation. Boosted algorithms such as gradient boost and logit boost have been used in literature to predict FFR and cardiovascular events (Weng et. al, 2017; Goldstein, 2017). Frizzell et al. and Mortazavi et al. built prediction models to determine the presence of cardiovascular disease using the 30-day readmission electronic data for patients with heart failure.

The reported C-statistic of the models varied from 0.533 to 0.628, showing an improvement in prediction with the machine learning approach over traditional statistical methods.

Numerous risk factor variables often make the prediction of CHD difficult, which in turn, increases the cost of diagnosis and treatment. In order to resolve the complexities and cost of diagnosis, advanced machine learning models are being widely used by researchers to predict CHD from clinical data of patients. Kurt et al. (2008) compared prediction performances of a number of machine learning models including the multilayer perceptron (MLP) and radial basis function (RBF) to predict the presence of CHD in 1245 subjects (Kurt et al., 2008). The MLP was found to be the most efficient method, yielding an area under the receiver operating characteristic (ROC) curve of 0.78. Shilaskar and Ghatol (2013) proposed a hybrid forward selection technique wherein they were able to select smaller subsets and increase the accuracy of the presence of cardiovascular disease with reduced number of attributes. Several other groups have reported techniques, such as artificial neural network (ANN), fuzzy logic (FL) and deep learning (DL) methods to improve heart disease diagnosis (Uyar, 2017; Venkatesh, 2017). Nonetheless, in most of the previous studies, the patient cohort was limited to a few thousand with limited risk factors.

We propose an efficient neural network with convolutional layers using the NHANES dataset to predict the occurrence of CHD. A complete set of clinical, laboratory and examination data are used in the analysis along with a feature selection technique by LASSO regression. Data preprocessing is performed using LASSO followed by a feature voting and elimination technique. The performance of the network is compared to several existing traditional ML models in conjunction with the identification of a set of important features for CHD. Our architecture is simple in design, elegant in concept, sophisticated in training schedule, effective in outcome with far-reaching applicability in problems with unbalanced datasets. Our research contributes to the existing studies in three primary ways: 1) our model uses a variable elimination technique using LASSO and feature voting as preprocessing steps; 2) we leverage a shallow neural network with convolutional layers, which improves CHD prediction rates compared to existing models with comparable subjects (the 'shallowness' is dictated by the scarcity of class-specific data to prevent overfitting of the network during training); 3) in conjunction with the architecture, we propose a simulated annealing-like training schedule that is shown to minimize the generalization error between train and test losses.

It is important to note that our work is not intended to provide a sophisticated architecture using a neural network. We also do not focus on providing theoretical explanation on how our network offers resistance to data imbalance. Instead, our goal is to establish that under certain constraints one can apply convolutional stages despite the scarcity of data and the absence of well-defined data augmentation techniques and to show that the shallow layers of convolution indeed offer resilience to the data imbalance problem by dint of a training schedule. The proposed pipeline contributes to improving CHD prediction rates in imbalanced clinical data, based on a robust feature selection technique using LASSO and shallow convolutional layers. This serves to improve prediction algorithms included in smart healthcare devices where sophisticated neural algorithms can learn from past user data to predict the probability of heart failure and strokes. Prediction rates could be integrated in healthcare analytics to provide real time monitoring which not only benefits the patients but also medical practitioners for efficient operations. The present research also focuses on a systematic training schedule which can be incorporated in smart devices to improve tracking of different predictor variable levels for heart failure. The rest of the paper is organized as follows: Section 2 explains data preparation and the preprocessing techniques. In section 3, we illustrate the convolutional neural network architecture with details on the training and testing methodology. In section 4 we demonstrate the results obtained from our model with performance evaluation metrics and compare it with existing models. Section 5 is the conclusion and discussion section. Here, several extensions to the research are proposed.

## 2. Data Preprocessing

Our study uses the NHANES data from 1999-2000 to 2015-2016. The dataset is compiled by combining the demographic, examination, laboratory and questionnaire data of 37,079 (CHD – 1300, Non-CHD – 35,779) individuals as shown in Figure 1. Demographic variables include age and gender of the survey participants at the time of screening. Participant weight, height, blood pressure and body mass index (BMI) from the examination data are also considered as a set of risk factor variables to study their effect on cardiovascular diseases. NHANES collects laboratory and survey data from participants once in every two years depending on their age and gender. In addition, based on the already existing validated experimental research, a comprehensive list of risk factor variables is selected from the laboratory tests conducted. Questionnaire data comprises of questions asked at home by interviewers using a Computer-Assisted Personal Interview (CAPI) system as mentioned in the NHANES website (NHANES, 2015). A total of 5 dichotomous predictor categorical variables are selected from the questionnaire data which have been shown to affect CHD (references, required). In all, 30 continuous and 6 categorical independent variables are used to predict the likelihood of coronary heart disease. For this study, coronary heart disease (CHD) is used as the dichotomous dependent variable. Awareness of CHD is defined as "yes" response to the question "Have you been ever told you had coronary heart disease?" Table I shows the categorical independent and dependent variables in the dataset considered for model development. The exhaustive list of variables is: gender, age, annual-family-income, ratio-family-income-poverty, 60sec pulse rate, systolic, diastolic, weight, height, body mass index, white blood cells, lymphocyte, monocyte, eosinophils, basophils, red blood cells, hemoglobin, mean cell volume, mean concentration of hemoglobin, platelet count, mean volume of platelets, neutrophils, hematocrit, red blood cell width, albumin, alkaline phosphatase (ALP), aspartate aminotransferase (AST), alanine aminotransferase (ALT), cholesterol, creatinine, glucose, gamma-glutamyl transferase (GGT), cholesterol, creatinine, glucose, iron, iron, lactate dehydrogenase (LDH), phosphorus, bilirubin, protein, uric acid, triglycerides, total cholesterol, high-density lipoprotein (HDL), glycohemoglobin, vigorous-work, moderate-work, health-Insurance, diabetes, blood related diabetes, and blood related stroke. However, in this list of variables, there are a couple of linearly dependent variables in terms of their nature of acquisition or quantification and some uncorrelated variables (annual family income, height, ratio of family income-poverty, 60 sec pulse rate, health insurance, lymphocyte, monocyte, eosinophils, total cholesterol, mean cell volume, mean concentration of hemoglobin, hematocrit, segmented neutrophils). We do not consider these variables for subsequent processing and analysis.

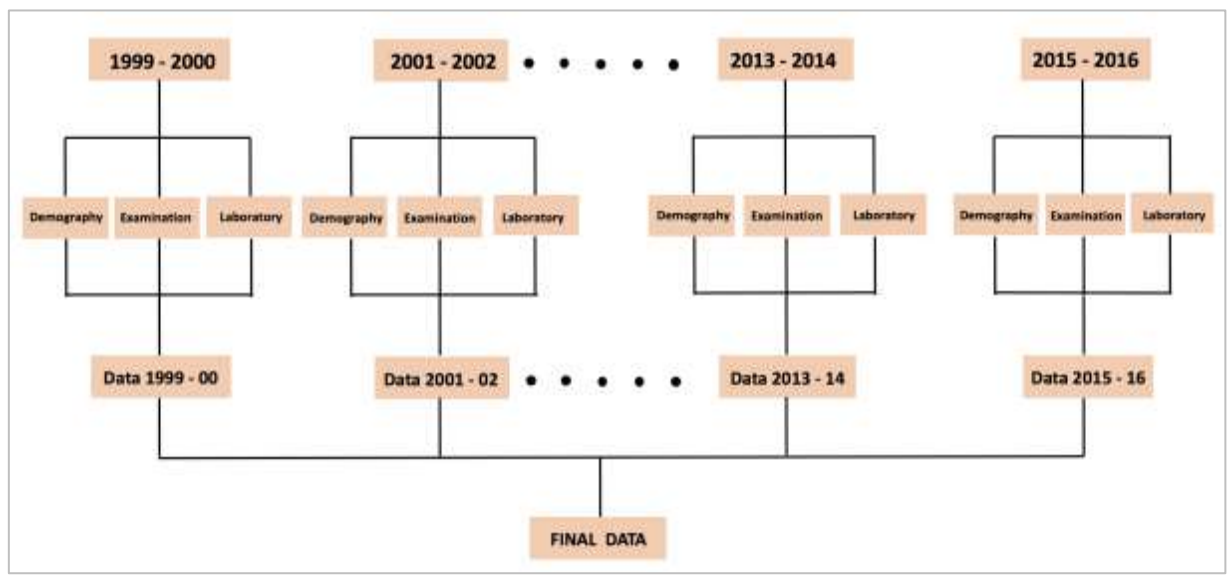

Figure 1. Data compilation from National Health and Nutritional Survey (NHANES). The data is acquired from 1999 to 2016 in three categories – Demography, Examination and Laboratory. Based on the nature of the factors that are considered, the dataset contains both the quantitative and the qualitative variables.

Table I: Description of the risk factor independent variables and the dependent variable.

| Variable Name | Description | Code | Meaning |
|---|---|---|---|
| Gender | Gender of the participant | 1<br>2 | Male<br>Female |
| Vigorous Activity<br>12 years and above | Vigorous activity in last one week or 30 days | 1<br>2<br>3 | Yes<br>No<br>Unable to do activity |
| Moderate Activity<br>12 years and above | Moderate activity in last one week or 30 days | 1<br>2<br>3 | Yes<br>No<br>Unable to do activity |
| Diabetes<br>1 yr and above | Doctor told that the participant has diabetes | 1<br>2<br>3 | Yes<br>No<br>Borderline |
| Blood Relative Diabetes<br>20 yrs and above | Biological blood relatives ever told that they have diabetes | 1<br>2 | Yes<br>No |
| Blood Relative Stroke<br>20 yrs and above | Biological blood relatives ever told that they have hypertension or stroke before the age of 50 | 1<br>2 | Yes<br>No |
| Coronary Heart Disease<br>20 yrs and above | Ever told that the participant had coronary heart disease | 1<br>2 | Yes<br>No |

## 3. Proposed Architecture

*3.1 LASSO Shrinkage and Majority Voting*

LASSO or least absolute shrinkage and selection operator is a regression technique for variable selection and regularization to enhance the prediction accuracy and interpretability of the statistical model it produces. In LASSO, data values are shrunk toward a central point, and this algorithm helps in variable selection and parameter elimination. This type of regression is well-suited for models with high multicollinearity. LASSO regression adds a penalty equal to the absolute value of the magnitude of coefficients, and some coefficients can become zero and are eventually eliminated from the model. This results in variable elimination and hence models with fewer coefficients. LASSO solutions are quadratic problems and the goal of the algorithm is to minimize:

$$\sum_{i=1}^{n} \left( y_i - \sum_j x_{ij}\gamma_j \right)^2 + \lambda \sum_{j=1}^{p} |\gamma_j| \qquad (1)$$

which is the same as minimizing the sum of squares with constraint $\sum |\gamma_j| \leq s$. Some of the $\gamma$ values are shrunk to exactly zero, resulting in a regression model that's easier to interpret. A tuning parameter, $\lambda$ which is the amount of shrinkage, controls the strength of the regularization penalty. When $\lambda = 0$, no parameters are eliminated. The estimate is equal to the one found with linear regression. As $\lambda$ increases, more coefficients are set to zero and eliminated. As $\lambda$ increases, bias increases and as $\lambda$ decreases, variance increases. The model intercept is usually left unchanged. The $\gamma$ value for a variable (factor) can be interpreted as the importance of the variable in terms of how the it contributes to the underlying variation in the data. The variable with a zero $\gamma$ is considered unimportant. It is to note that LASSO shows misleading results in case of data imbalance, which may prompt incorrect selection of important variables if we perform LASSO on the entire dataset.

Note that the variables in our dataset are mixed data type – a subset of them are categorical. In this work, as a standard practice, we use group LASSO and we refer it as LASSO for simplicity. In order to mitigate the effect of imbalance, we adopt a strategy to randomly subsample the dataset and iterate LASSO multiple times. Majority voting is performed on the set of $\gamma$ values to identify the variable that are nonzero in major number of iterations. Let us assume, that LASSO is performed N times on N randomly subsampled dataset, where each instance has equal number of examples in case of CHD and no-CHD. With 45 variables at hand, we obtain $\gamma_i = [\gamma_{i,1}\ \gamma_{i,2} \ldots \ldots \gamma_{i,45}]$ at $i^{th}$ instance of LASSO. For any variable $c$, we count the number of instances in which the variable is non-zero, and with a manually set threshold, we decide the selection of that variable for further analysis. Mathematically,

$$\chi(\gamma) = \begin{cases} 1 & if\ \gamma \neq 0 \\ 0 & otherwise \end{cases}$$

$$[\chi(\gamma_{1,c})\ \chi(\gamma_{2,c})\ \ldots\ldots\ \chi(\gamma_{N,c})]\mathbf{1} \geq \frac{N}{\alpha} \Rightarrow \mathbf{c}\ \text{is selected} \tag{2}$$

*3.2. Convolutional Neural Network (CNN)*

The challenge of predicting the existence of CHD in patients refers to the task of binary classification. Under certain constraints, neural network (LeCun, 2015; Goodfellow, 2016) has been proven to be an effective parametric classifier under supervised settings. Recently, with the explosion of structured data, deep neural networks incorporating a large number of application-specific hidden layers, have demonstrated significant improvement in several areas including speech processing, applications involving image processing, and time series prediction (LeCun, 1995). There is a vast body of deep learning architectures that are fine-tuned and rigorously trained using big datasets. An artificial neural network (Iandola, 2016; Krizhevsky, 2012; Szegedy, 2017; He, 2016) successively transforms the input data over sequential hidden layers and estimates the error at the output layer. The error is back propagated to iteratively update the layer weights using gradient descent algorithm. Rigorous experimentations and analyses have proposed several improvements in the gradient descent algorithm, the nonlinearity of layers, overfitting reduction, training schedule, hidden layer visualization and other modifications. Despite resounding success in applications, the working principle of deep neural networks is still poorly understood. It is also found in practice that a deep neural network is extremely susceptible to be attacked by adversarial examples. In addition, owing to millions of parameters in a typical *deep* architecture, the trained network may be overfitted, especially in cases where there is scarcity of examples.

Among various algorithms that attempted to overcome this problem, data augmentation (Krizhevsky, 2012; Radford, 2015) is a widely used technique that artificially generates examples to populate small datasets. However, such a procedure is biologically implausible in most clinical datasets. For example, augmented measurements of a CHD phenotype, such as platelet count, might not correspond to possible readings of a subject. It is because the underlying principles of the statistical generation and the biological sources of platelet count readings may be fundamentally different. Poor training due to small or imbalanced datasets and susceptibility to adversarial examples lead to poor and unreasonable classification. Unlike many computer vision tasks, such as semantic labeling, chat-bot configuration, and hallucinogenic image synthesis

(Mordvintsev, 2015), erroneous prediction in medical research is accompanied by a significant penalty.

For example, faulty prediction of a subject having chronic CHD may leave the subject untreated or misdirect the possible therapeutic medication. Therefore, one of the prime objectives of this paper is to improve classification accuracy, i.e. the prediction accuracy of the subjects with and without the presence of CHD. There are several other relevant concerns related to misclassification in medical research (Marcus, 2018). To overcome these limitations and driven by the success of deep networks, we propose a shallow convolutional neural network, where the convolution layers are 'sandwiched' between two fully connected layers as shown in Fig. 2.

*3.2. CNN Architecture*

The architecture is a sequential one-input-one-output feedforward network. For simplicity, we assume the class of subjects with presence of CHD as class '1' and the subjects with absence of CHD as class '0'. As mentioned in the previous section, the number of active phenotypes of CHD obtained from majority voting is 50. Let the number of training examples be *N*, which indicates that the input layer in Fig. 2 has dimension of $\mathcal{R}^{N \times 50}$. The dense or fully connected layers, consisting of 64 neurons collectively act as a linear combiner of the 50 variables and bias, which effectively homogenizes different variable types before nonlinear transformation. The nonlinear transformation is carried out by rectified linear unit (ReLU). Dropout with 20% probability is performed to reduce overfitting. Following the fully connected layer, there is a cascade of convolution layers. In the first convolution layer, there are two filters of kernel width 3 and stride 1. The layer is not provided with external zero-padding. In the pooling layer, we rigorously experiment with different pooling strategies and find average pooling working marginally better than max pooling under all constraints. The first convolution layer converts the output of fully connected block $\in \mathcal{R}^{N \times 64}$ to a tensor of dimension $\mathcal{R}^{N \times 64 \times 1}$. The tensor is then subjected to batch normalization, nonlinear transformation and average pooling with an output tensor of dimension $\mathcal{R}^{N \times 31 \times 2}$.

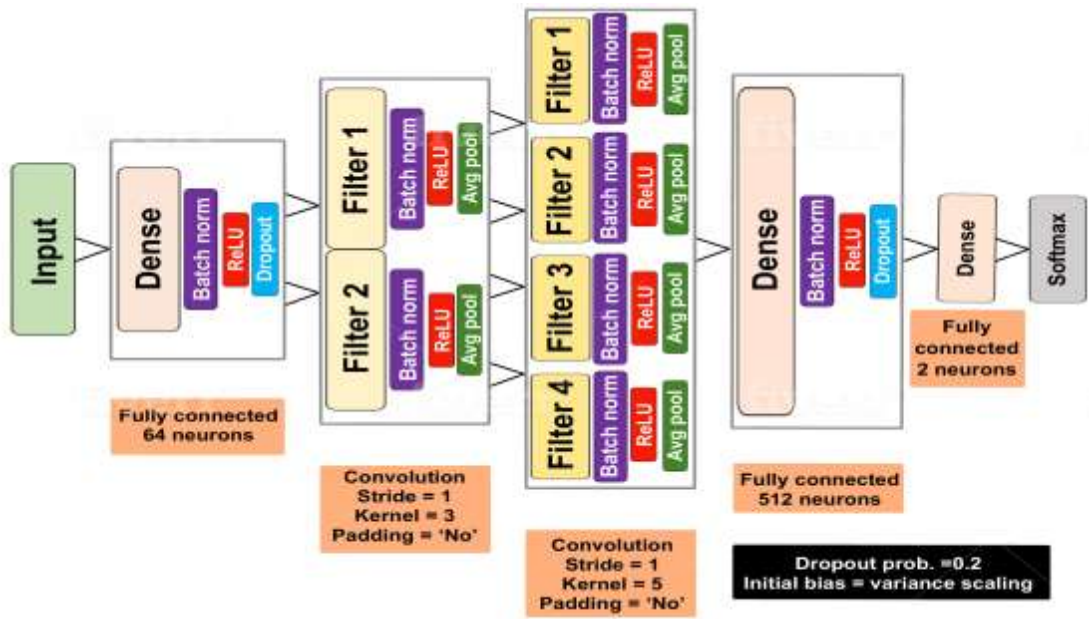

Fig 2. Proposed convolutional neural network architecture. The 'Input' is a 1D numerical array corresponding to all the factors/variables from LASSO-Majority Voting preprocessing stage. The 'Dense' layer, immediately after the 'Input', combines all the factors and each neuron (computing node) at the output of 'Dense' layer is a weighted combination of all the variables, indicating a homogeneous mix of different variable types. The next two convolution layers seek representation of the input variables via the 'Dense' layer. The next two 'Dense' layers are followed by the 'Softmax' layer. The last two 'Dense' layers (before the 'Softmax' layer) can be retrained for transfer learning in case new data is obtained. The associated training parameters, such as dropout probability, number of neurons, activation function (we used ReLU), pooling types, and convolution filter parameters are shown in the above figure. Owing to the large number of parameters that can lead to overfitting of training data points, we propose a training schedule in section 3.2.1.

The filters in the last no-zero-padded convolution layers are taken with kernel 5 and stride 1, delivering an output tensor of $\mathcal{R}^{N \times 13 \times 4}$ to the next dense layer after the average pooling layers. The categorical output is observed at the end of the softmax layer, where we set the loss function as the categorical cross-entropy loss. The bias in each layer is initialized with random numbers drawn from a truncated normal distribution with variance $\frac{1}{\sqrt{n}}$, where $n$ is the number of 'fan-in' connections to the layer. We use Adam optimizer with learning rate 0.005, $\beta_1 = 0.9$, $\beta_2 = 0.999$ and zero decay. Our proposed architecture consists of 32,642 trainable and 1,164 non-trainable parameters. We experiment with several hyperparameters that are associated with our model to obtain consistent class-wise accuracy. We provide results by varying subsampling of input data, epochs, class-weights, the number of neurons in each dense layer except the last one, and the number of filters in each convolution layer during training.

*3.2.1. Training schedule*

During training, the class weight ratio, which is adjusted as a penalty factor due to class imbalance, is defined as the ratio of CHD and Non-CHD dataset. For example, a class weight ratio of 10:1 indicates that any misclassification of a CHD training sample will be penalized 10 times more than a misclassified Non-CHD sample during the error calculation at the output prior to backpropagation after each epoch. Although, we use dropout layers in our CNN model, we also use this training schedule in order to further reduce possible overfitting. The intuition is to initially train the model with 1:N weight ratio for sufficiently large number of epochs and then, gradually increase the weight ratio with a steady decline in the number of epochs. Let the actual class weight ratio is $\rho_0: 1$, which we take as a factor $\rho_0$.

---

Fitting our CNN model, M, by varying the number of epochs ($\omega$) and weight ratio ($\rho$)

1. Initialize $\rho$ = N, $\omega$ (large number, we set as N), M, end_iter (5-10 depending on the instance), i=1
2. While $\rho \leq \rho_0$
    M.fit (Data, $\omega, \rho$)
    $\rho \leftarrow floor(\frac{\rho}{2})$
    $\omega \leftarrow floor\left(\frac{\omega}{2}\right)$
3. While $(i \leq end\_iter)$ and $Trainloss(i) \leq Trainloss(i-1)$
    M.fit (Data, $i, \rho_0$)

---

*3.3 Competitive Approaches*

Machine learning classification methods have shown to potentially improve prediction outcomes in coronary heart disease. Such classification methods include logistic regression, support vector machines, random forests, boosting methods and multilayer perceptron (Goldstein, 2017). Logistic regression models the prediction of a binomial outcome with one or more explanatory variables, using a standard logistic function which measures the relationship between the categorical dependent variable and one or more independent variables by estimating the probabilities. The logistic function is given by, $f(x) = \frac{1}{1+e^{-x}}$ which in common practice is known as the sigmoid curve. Support vector machine (SVM) is a binary classification algorithm which generates a (*N*-1) dimensional hyperplane to separate points into two distinct classes in an *N* dimensional plane. The classification hyperplane is constructed in a high dimensional space that represents the largest separation, or margin, between the two classes.

Random forests are an ensemble learning algorithm where decision trees that grow deep are averaged and trained on different parts of the training set to reduce variance and avoid overfitting. Random forests algorithm employs bagging or bootstrap aggregating and at each split a random subset of features are selected. Bagging is a parallel ensemble because each model is built independently. Boosting on the other hand is a sequential ensemble where each model is built based on correcting the misclassifications of the previous model. In boosting methods, the weights are initialized on training samples and for *n* iterations, a classifier is trained using a single feature and training error evaluated. Then the classifier with the lowest error is chosen and the weights are updated accordingly; the final classifier is formed as a linear combination of *n* classifiers. A boost classifier is in the form, $F_T(x) = \sum_{t=1}^{T} f_t(x)$ where each $f_t$ is a weak learner with $x$ as input. Each weak learner produces an output hypothesis, $h(x_i)$, for each sample in the training set. At each iteration *t,* a weak learner is selected and assigned a coefficient $\alpha_t$ such that the sum training error $E_t$ of the resulting *t*-stage boost classifier is minimized. A multilayer perceptron (MLP) is a feedforward artificial neural network (ANN) which consists of an input layer, an output layer and one or more hidden layers and utilizes backpropagation for training the data. The MLP commonly uses a nonlinear activation function which maps the weighted inputs to the output of each neurons in the hidden layers. In an MLP, the connection weights are changed based on the error between the generated output and expected result. Two of the most common activation functions are the rectified linear unit (ReLU), $f(x) = x^+$ and the hyperbolic tangent, $y(x_i) = \tanh(x_i)$.

Data augmentation demands attention in the context of data imbalance. Algorithms, such as random oversampling (ROS), synthetic minority over-sampling technique (SMOTE) (Chawla, 2002), and adaptive synthetic sampling (ADASYN) (He, 2008) augment the minority class data by either replicating or synthesizing new data. One pertinent issue with regard to synthesized data is that, unlike images, the data in the context of biological factors (variables) may be implausible as it is difficult to verify the authenticity of newly augmented data, especially when both classes of data are closely spaced. In this paper, we provide comparative results by using the above algorithms. In addition, we provide the corresponding visualization of the augmented data via t-SNE. Please keep in mind that t-SNE is a non-convex algorithm, generating embedding that depends on the initialization low-dimensional embedding. We employ random undersampling strategy to select a subset of data for training the CNN. Similar to data augmentation, there are several data undersampling strategies. We compare our results with edited nearest neighbor (EDN) (Wilson, 1972), instance hardness threshold (IHT) (Smith *et al.*, 2014) and three versions of near-miss (NM-v1, v2 and v3) (Mani, 2003) algorithms.

## 4 Results

*4.1 Summary Statistics*

For the purpose of variable selection in our classification model, we start out by investigating correlations among 30 continuous predictor variables. Correlation is found to be high (0.77) between serum alanine aminotransferase (ALT) and aspartate aminotransferase (AST) as given from Fig 3. However, AST is a major risk factor in the prediction of CHD, as has been reported in the literature (Shen et. al, 2015). Jianying (2015) finds AST levels to be significantly higher in CHD patients than in the control group and hence can be used as biochemical markers to predict the severity of CHD. A high correlation of 0.89 was determined between body-mass-index and weight which seemed normal whereas correlation between hemoglobin and red blood cells was 0.74. While the association of hemoglobin with clinically recognized CHD is limited in research (Chonchol, 2008), the role of red blood cells in CHD is well researched in the literature (Madjid, 2013). It has been investigated that high blood glucose levels for non-diabetic patients can significantly increase the risk for development of CHD (Neilson, 2006) which is depicted in Fig. 3 by a high correlation value of 0.79 between glycohemoglobin and glucose. Further, we find a correlation of 0.46 between protein and albumin. Lower levels of serum albumin have been reported to be linked with increased levels of cardiovascular mortality as well as CHD (Shaper, 2004) while higher level of protein is reported to increase risks of CHD (Clifton, 2011). Serum Lactate dehydrogenase (LDH) is found to be correlated with AST (correlation coefficient of 0.41), consistent with previous studies which suggest that increased value of LDH in active population is associated with low risk of CHD (Kopel et. al., 2012). Due to the importance of the risk factors (as reported in existing literature) of some of the correlated variables and their association with CHD, LASSO regression was performed to correctly determine the predictor variables for further analysis.

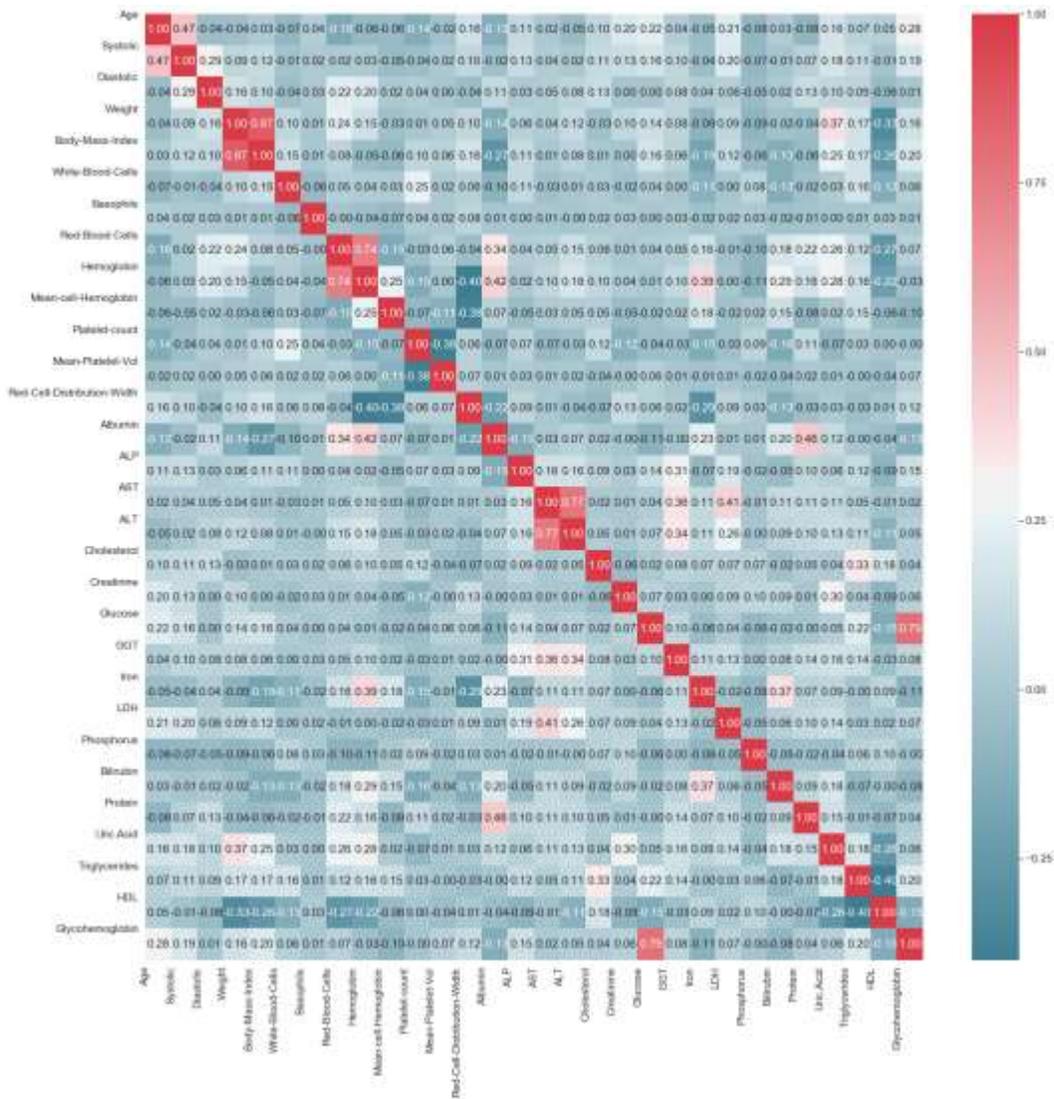

Fig. 3. Correlation table for the independent predictor variables. In this table, moderately strong correlations among few pairs are observed (Glucose and Glycohemoglobin, Red blood cells and Hemoglobin, ALT and AST, Weight and Body-Mass-Index). Rest of the pairs show fairly low correlation values, implying the variables after the LASSO-Majority voting stage are sufficiently decorrelated.

*4.2 Model results*

To identify variables that contribute to the variation in data, we apply LASSO to 100 instances of randomly sampled datasets, with each containing 1300 examples of class CHD (negative class) and 1300 of class no-CHD (positive class). We set $\alpha$ in eq. (2) as 6 and find that ALT, glucose, hemoglobin, body mass index fails to contribute significantly in the data irrespective of strong experimental evidence in state-of-the-arts that favor those factors. From majority voting, some of the strong correlates are age, white blood cells, platelet count, red cell distribution width, cholesterol, LDH, uric acid, triglycerides, HDL, glycohemoglobin, gender, presence of diabetes, blood related stroke, moderate and vigorous work.

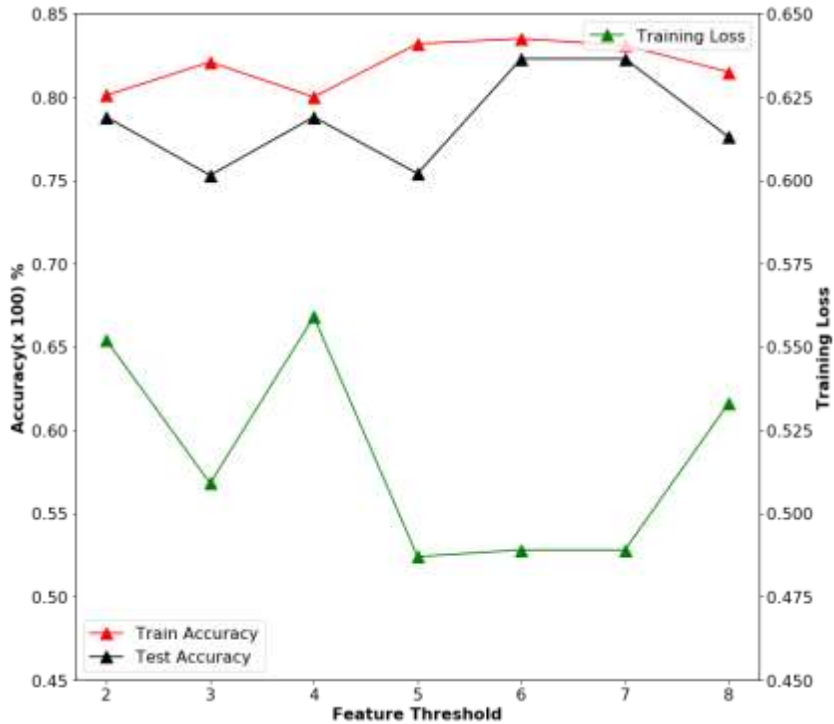

Fig. 4. Model accuracy as a function of majority voting threshold. The threshold value of majority voting affects the classification accuracy of CHD as the selection of this value controls the number of variables that are to be channeled to our CNN model. The smaller is the threshold value, the larger is the set of variables. Based on the training loss, training accuracy and test accuracy, the threshold value between 16.67 (100/6) – 20 (100/5) combining 100 instances of LASSO appears suitable for obtaining balanced per class (CHD and Non-CHD) classification accuracy.

To achieve the optimal number of features for training our CNN architecture, the threshold of the feature voting was kept in the range of 2 to 8. The highest accuracy obtained from training the network is 83.17% with a training loss of 0.489 with a threshold feature of 6 as shown in Fig. 4. The corresponding highest test accuracy obtained is 82.32%. LASSO regularization reduces the coefficients of three of the continuous predictor variables (Body-mass-index, glucose and ALT) and one categorical variable to zero which is determined to be highly correlated as seen in section 4.1. With a threshold value of 6, the CNN architecture is trained separately on different sets of subsampled data sets. Subsampling is performed in varying ratios starting with the range 1300:13000 (CHD: Non-CHD) and increased to 1300:4000 as shown in Table II. The corresponding test accuracy is reported as 82.32%.

Table II: Training schedule for increasing class weight ratio and sampling for optimal threshold. A maximum training accuracy if 83.51% and minimum training loss of 0.489 is achieved with a misclassification penalty of 3:1 (CHD: Non-CHD) and a sampling ratio of 1300:4000 (CHD: Non-CHD).

| Class Weight | Sampling | TPR | TNR | Train Accuracy (%) | Test Accuracy (%) | Training Loss |
|---|---|---|---|---|---|---|
| 10:1 | 1300:13000 | 0.690 | 0.812 | 78.00 | 81.09 | 0.684 |
| 8:1 | 1300:10000 | 0.706 | 0.827 | 77.28 | 82.67 | 0.676 |
| 6:1 | 1300:8000 | 0.741 | 0.799 | 80.00 | 79.90 | 0.595 |
| 5:1 | 1300:6000 | 0.735 | 0.80 | 80.48 | 79.96 | 0.548 |
| 3:1 | 1300:4000 | 0.778 | 0.823 | 83.51 | 82.32 | 0.489 |

In each subsampled set, all CHD subjects are taken into consideration and the Non-CHD subjects are randomly chosen without replacement. For training purposes, the neural network is trained by varying the misclassification penalty from 10:1 (CHD: Non-CHD) to 3:1 as shown in Table II. The maximum accuracy of 83.51% and minimum training loss of 0.489 is obtained while training with a sampling ratio of 1300:4000 (CHD: Non-CHD) and misclassification penalty of 3:1(CHD: Non-CHD). The trained network is tested on a cohort of 31,779 (85.70% of the whole dataset) remaining samples and a test accuracy of 82.32% is obtained subsequently as reported in table II. The optimal sampling ratio (1300:4000) of class 1 to class 0 is maintained for the final training of the network as illustrated in Table III.

Table III reports the final training of the CNN architecture with varying class (CHD: Non-CHD) weights to check for consistency of our results. With decrease in class weights, the training accuracy increases from 59.43% to 83.17% when the difference between the training and test accuracies becomes a minimum, indicating a reasonable generalization error as shown in Fig. 5. During training the number of epochs, the number of neurons in each dense layer except the last one, and the number of filters in each convolution layer are varied to obtain the best fit of the model. We have fine-tuned several hyperparameters that are associated with our model to obtain consistent class-wise accuracy. The optimization is performed using Adam as the activation function with a learning rate 0.006 and 60 epochs, and no scheduling of learning rate is used during the training. The best test accuracy obtained is 82.32% where the penalty of misclassification of class 1 is set three times higher than that of class 0.

Table III: Training schedule for increasing class weight ratio and optimal sampling ratio attained from table I. The difference between the training (83.17%) and test (82.32%) accuracies attain the minimum when the misclassification penalty of class 1 is set three times higher than class 0.

| Class Weight | TPR | TNR | Train Accuracy (%) | Test Accuracy (%) | Training Loss |
|:---:|:---:|:---:|:---:|:---:|:---:|
| 50:1 | 0.980 | 0.383 | 59.43 | 44.48 | 1.591 |
| 25:1 | 0.923 | 0.568 | 67.06 | 58.04 | 1.236 |
| 12:1 | 0.860 | 0.664 | 71.98 | 66.60 | 0.965 |
| 8:1 | 0.836 | 0.686 | 75.08 | 69.76 | 0.765 |
| 6:1 | 0.817 | 0.740 | 80.13 | 74.12 | 0.620 |
| 4:1 | 0.788 | 0.770 | 81.00 | 77.10 | 0.550 |
| 3:1 | 0.773 | 0.818 | 83.17 | 82.32 | 0.489 |

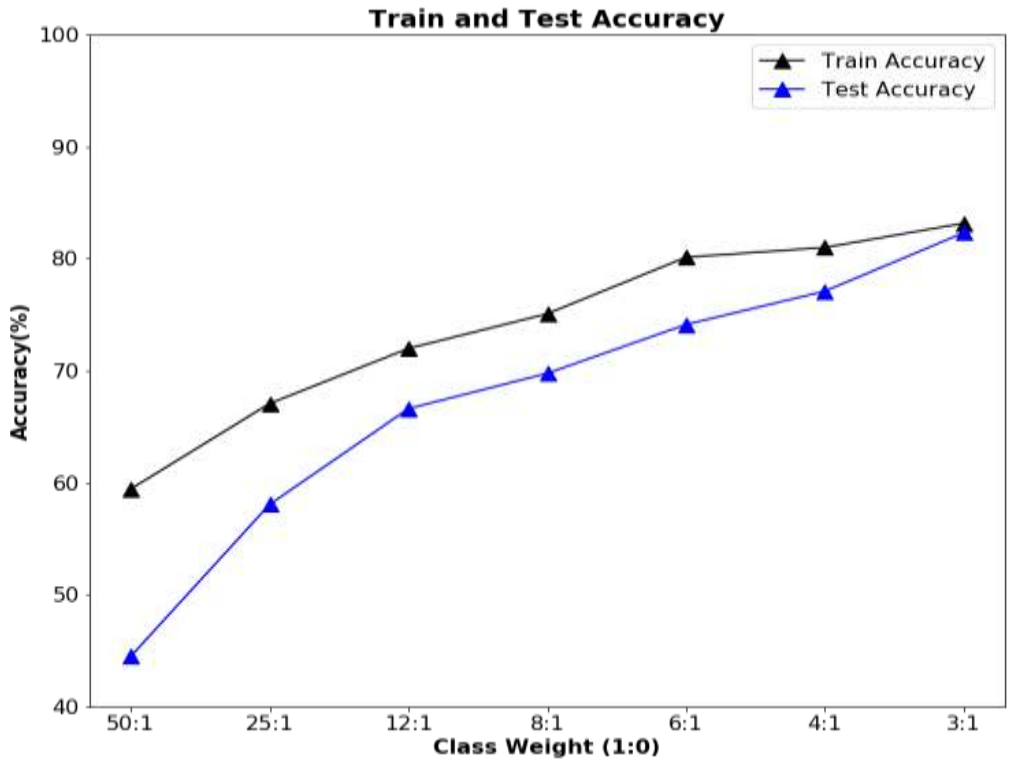

Fig. 5. Training and test accuracies with varying misclassification penalties for class 1 and 0. The minimum difference between training and test accuracies is obtained with a class weight of 3:1 (CHD: Non-CHD) and a training loss = 0.489. The model is trained with a constant optimized learning rate of 0.006 and 60 epochs.

The performance of the proposed CNN classifier can be evaluated from the confusion matrix in Table IV. We specify the classification parameters as follows:

TP: true positive classification cases (true predictions for class 1, i.e., true CHD predictions),

TN: true negative classification cases (true predictions for class 0, i.e., true non- CHD predictions),

FN: false negative classification cases (false predictions for class 1, i.e., false CHD predictions),

FP: false positive classification cases (false predictions for class 0, i.e., false non- CHD predictions).

Some commonly applied performance rates calculated are the true positive rate (TPR), the accuracy of predicting CHD (class 1) and the true negative rate (TNR), the accuracy of predicting non-CHD (class 0). The detailed values of TPR, TNR, train accuracy, test accuracy and training loss for all class weights are given in Table III.

Table IV: Confusion matrix for the CNN classifier for coronary heart disease. Out of 208 coronary heart disease cases in the sample cohort, 161 cases were predicted correctly by the classifier. The proposed classifier also correctly predicts 25,828 cases where patient did not report coronary heart disease.

| Total Cohort | | True Condition | |
|---|---|---|---|
| | | Presence of CHD | Absence of CHD |
| **Predicted Condition** | Presence CHD | True Positive (TP) = 161 | False Negative (FN) = 47 |
| | Absence CHD | False Positive (FP) = 5743 | True Negative (TN) = 25828 |

In the present study, it is our objective for our classifier to predict the presence of CHD with higher (improved) accuracy than in previous studies. The recall rate (sensitivity) for correctly predicting the true positive rate for class 1 (presence of CHD) is 77% while the class 0 (absence of CHD) is 81%. The CNN classifier has been tested on 31,779 subjects while maintaining almost the same TPR and TNR, contrary to previous reported studies which have considered significantly smaller samples.

Partitioning a highly imbalanced dataset poses a lot of challenges and incurs unavoidable biases on a classifier's performance. While, it is advised to keep same ratio of class-specific data in training and testing in state of the arts, however, in a highly imbalanced dataset it is often very challenging to get a high class 1 (true CHD prediction rate) accuracy as the testing data size increases. The present method confirms that our proposed CNN architecture has the classification power of 77% to correctly classify the presence of CHD cases on a testing data, which is 85.70% of the total dataset. This result signifies that the proposed architecture can be generalized to other studies in healthcare with a similar order of features and imbalances.

The performance of our binary classifier is calculated by computing the ROC curve (Yang et. al; 2017). The area under the curve (AUC) value in the ROC curve is the probability that our proposed

CNN classifier ranks a randomly chosen positive case (CHD) higher than a randomly chosen negative case (Non-CHD) (Tom 2005). Thus, the ROC curve behaves as a tool to select the possible optimal models and to reject the suboptimal ones, independently from the class distribution. It does so by plotting parametrically the true positive rate (TPR) vs the false positive rate (FPR) at various threshold settings as shown in Fig. 4. (Right). The calculated AUC is 0.767 or 76.7% which is comparable to previous studies related to CHD (Martinez, Schwarcz, Valdez & Diaz, 2018). In highly imbalanced data sets balanced accuracy is often considered to be a more accurate metric than normal accuracy itself. The balanced accuracy of the model is determined to be (TPR + TNR)/2 = 0.795 or 79.5%. The fall out rate or the Type-I error of the model is 5743/31571 = 18.2% and the miss rate or the Type-II error of the model is 47/210 = 22.6%. The positive likelihood ratio of the predicted model is 4.27 indicating that there is almost a 30% increase in probability post diagnosis in prediction of the presence of CHD in patients. A negative likelihood ratio of 0.27 was calculated which signifies that there is approximately 30% decrease in probability post diagnosis in prediction of absence of CHD in patients.

*4.3. Comparison of ML models*

*4.3.1. Comparison with state-of-the-art ML models*

Machine learning models discussed in section 3.3 are implemented and tested on our test cohort. The prediction results from these methods are then compared with the results of our proposed CNN architecture. All models are implemented with optimized parameters and then compared based on the true positive rate (recall) and true negative rate (sensitivity). Corresponding test accuracies and AUC values are also determined. Logistic regression and adaboost classification result in highest test accuracies but these classifiers suffer from low recall values which is the true positive rate for coronary heart disease detection. While the recall values obtained from SVM and random forest are comparable to that of our proposed CNN model, our model predicts the negative (Non-CHD) cases with higher accuracy as shown in Table V. The balanced accuracy of our model (79.5%) is also higher than individual accuracies from that of SVM or random forest classifiers. An optimized two-layer multilayer perceptron resulted in a low recall value of 66.34% when tested on our test cohort. Results in Table V show that SVM and random forest classifiers perform better than logistic, adaboost and MLP classifiers, but the specificity and test accuracy are significantly lower as compared to our designed CNN classifier. The CNN classifier results in high specificity and test accuracy along with high values of recall and AUC.

Table V: Comparison of machine learning models for coronary heart disease prediction. As compared to traditional machine learning models, our proposed model attains a recall value of 0.77 which is comparable to the SVM classifier. However, specificity (0.81) and test accuracy (0.82) of our model are significantly higher than the SVM classifier.

|  | Recall (%) | Specificity (%) | Test Accuracy (%) | AUC |
| --- | --- | --- | --- | --- |
| Logistic Regression | 51.44 | 91.15 | 90.89 | 71.29 |
| SVM | 77.40 | 77.87 | 77.87 | 77.64 |
| Random Forest | 76.44 | 76.06 | 76.06 | 76.25 |
| AdaBoost | 52.88 | 90.36 | 90.12 | 71.63 |
| MLP | 66.34 | 78.88 | 78.80 | 72.61 |
| Our model | 77.3 | 81.8 | 81.78 | 76.78 |

These results confirm that the CNN classifier outperforms all existing commonly used machine learning models for coronary heart disease prediction in terms of accuracy in prediction for both CHD and Non-CHD classes.

*4.3.2. Our LASSO-CNN vs vanilla CNN*

Our experimental set up is generalized in the sense that setting $\frac{N}{\alpha}$ as 0 (or, equivalently $\alpha = \infty$) will select all the variables into consideration. Thus, applying vanilla CNN is equivalent to applying the model LASSO ($\infty$) - CNN. Using the same subsampled dataset with 3:1 ratio of class samples, vanilla CNN yields 79.42% test accuracy, which is approximately 2% less than the average test accuracy that we obtain on an average by applying LASSO (6)-CNN. Although, it seems a marginal improvement, the number of samples in excess that are accurately labeled by our model is 635 ($\approx$31779*0.02).

*4.3.3. Data oversampling strategies*

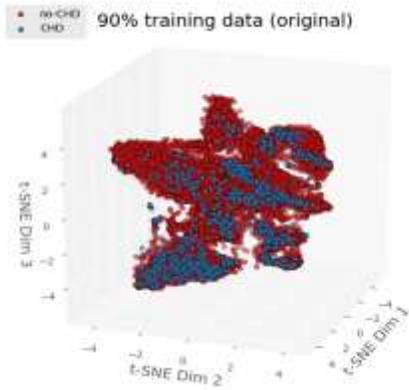
(a)

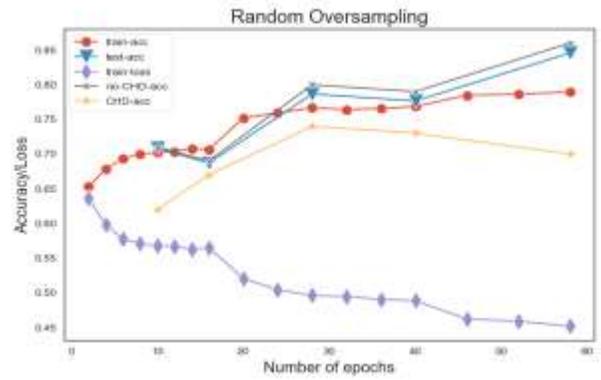
(b)

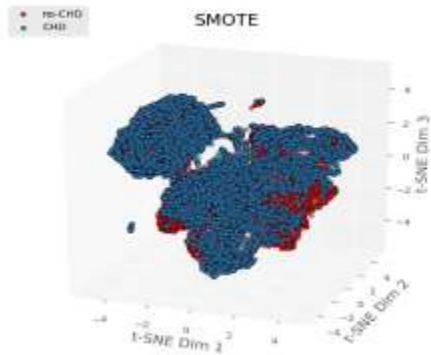
(c)

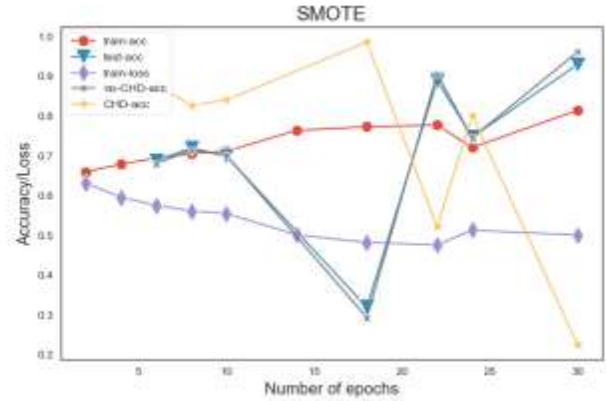
(d)

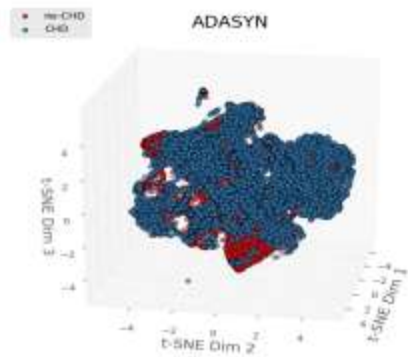
(e)

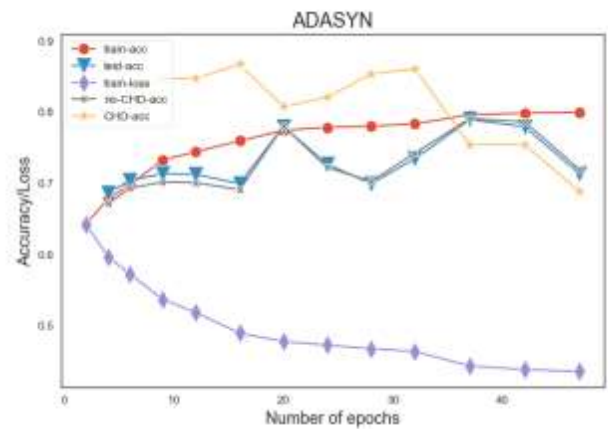
(f)

Fig. 6. Results using three oversampling techniques for the data augmentation of the minority class. For each technique, the results provide training accuracy, test accuracy, training loss, CHD accuracy (class-specific) and no-CHD accuracy (class-specific) over a number of epochs, added with t-SNE low-

dimensional embedding for data visualization in 3D. (a) t-SNE visualization of 90% of the original data used for training. 10% of the data is reserved for testing. (b) The results using random oversampling (ROS). Note that we did not provide the t-SNE visualization for ROS as, in ROS, data samples from the minority class are randomly picked and added to the data, thereby maintaining the same data with redundant samples. So, the visualization is same as the original data in (a). (c)-(d) Results using SMOTE with visualization. (e)-(f) Results using ADASYN with visualization.

We compare our LASSO-CNN with the state-of-the-art oversampling algorithms. Note that the size of the testing dataset is remarkably small in oversampling cases, where the minority class is oversampled to become cardinally same with the majority class. In case of ROS and SMOTE, each class of the training data contains 32013 samples and the test data has 3709 samples (CHD: 3558 samples, no-CHD: 151 samples). In case of ADASYN, the training data size is 64134 (CHD: 32121 samples, no-CHD: 32013 samples) and the test data size is identical to ROS.

In all the cases, the ratio of class-specific samples (no-CHD:CHD) in the test data is approximately 24:1. From Fig. 6, it can be seen that, in all the three algorithms, the test accuracy of the no-CHD class strongly follows the overall test accuracy, indicating the predominant effect of data imbalance in the test data. Except in ADASYN, it is observed that the test accuracy increases with our model being trained using more epochs. The increase in the test accuracy is favored by the increase in the no-CHD accuracy at the expense of compromising CHD accuracy.

Overall, the behavior of LASSO+CNN using SMOTE is erratic, whereas trends in train accuracy, test accuracy, train loss can be clearly observed in cases of ROS and ADASYN. Although, the test accuracy is higher in case of SMOTE when compared with ROS and ADASYN, the class-specific accuracies are balanced in case of ADASYN, yielding 79%, 79.14% and 75.5% for test accuracy, no-CHD accuracy and CHD accuracy respectively, with the class score variance of 3.06. In terms of class score variance, ROS trails behind ADASYN, scoring 78.67%, 80% and 74% with 9 as the class score variance. Class score variance measures how close the CHD and no-CHD accuracies are.

*4.3.4. Data undersampling strategies*

We train and test our LASSO+CNN model on several datasets that are undersampled by NM (v1, v2 and v3), IHT and EDN, as discussed in section 3.3. In case of undersampled datasets, the size of the test dataset is large as the size of the majority class (no-CHD) is reduced against the minority class (CHD). After applying an undersampling technique, the samples that are not considered in

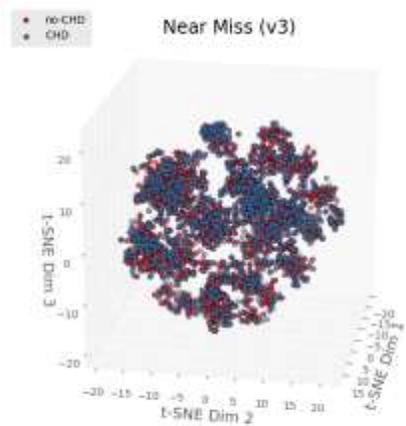

(a)

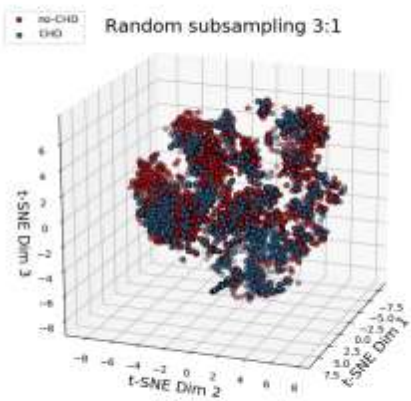

(b)

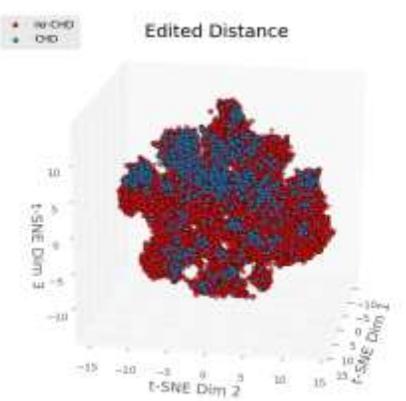

(c)

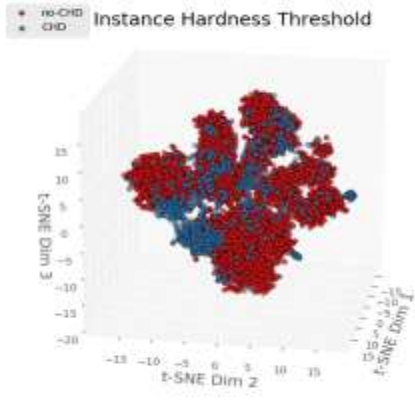

(d)

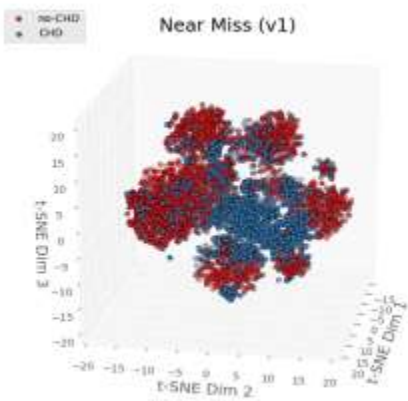

(e)

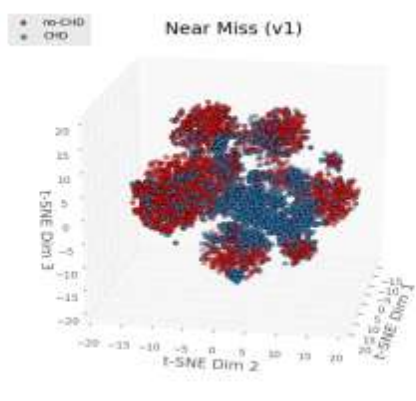

(f)

Fig. 7. t-SNE visualization of the five undersampling techniques for the data reduction of the majority class. (a), (e) and (f) Near-miss using k-nearest neighbor (version-1,2 and 3). (b) Random subsampling with 3:1 as no-CHD: CHD data samples (one instance). (c) Edited nearest neighbor (EDN). (d) Instance hardness threshold (IHT).

the train dataset is added to the test dataset. In case of EDN, the test data contains 6376 samples (CHD: 152 samples, no-CHD: 6224 samples), whereas in case of IHT, the test data has 24995 samples (CHD: 152 samples, no-CHD: 24843). Three versions of near-miss (NM) are employed. Each of version1and 2 contains 31200 data samples (CHD: 152 samples, no-CHD: 31048 samples), where version 3 has 32557 samples (CHD: 152 samples, no-CHD: 32405 samples). In all of NM versions, there are 4523 no-CHD and 1357 CHD training data, maintaining approximately 3.5:1 data ratio to enforce similar experimental set up where we report the best test accuracy of 82.32% by our model in Table II.

Fig. 7 provides t-SNE visualization of undersampled data using the algorithms mentioned above. It is evident that random subsampling and near-miss (v3) encompass the span of the data uniformly and better than other algorithms that we consider. In fact, near-miss (v3) yields 74% no-CHD accuracy and 87% CHD accuracy, with the overall test accuracy of 74.59%. Fig. 8 depicts the class wise accuracies by the undersampling methods on the test data. Although the test data size majorly differs in over- and undersampled data, note that the no-CHD accuracy significantly drops in case of undersampling except in random undersampling (Table IV). No-CHD is the majority class and the reason behind such accuracy drop might be two-fold. Firstly, the reduced data size insufficiently capture the true population variation, thereby failing to generalize on the test data. This is true for all the undersampling strategies. The second reason depends on the constraint of the algorithm. For example, to obtain 1:1 class data for training, near-miss (v3) first selects m neighbors for every minority class sample, collectively formed the set S. Later, among the majority-class samples in S, the algorithm selects a subset, where each majority-class sample has the largest k nearest neighbor. The "largest" distance ensures for the maximal separation (local margin) of classes, foreshadowing a potential loophole for the classification of majority samples which fall in between. Random subsampling does not have such constraints. A better strategy would be, at first, the selection of data ratio, which we found as 3:1 (no-CHD: CHD). This value depends on the dataset. Later, an ensemble of LASSO+CNN classifiers can be built for a set of randomly subsampled datasets. Testing of an unknown sample can be performed using majority voting.

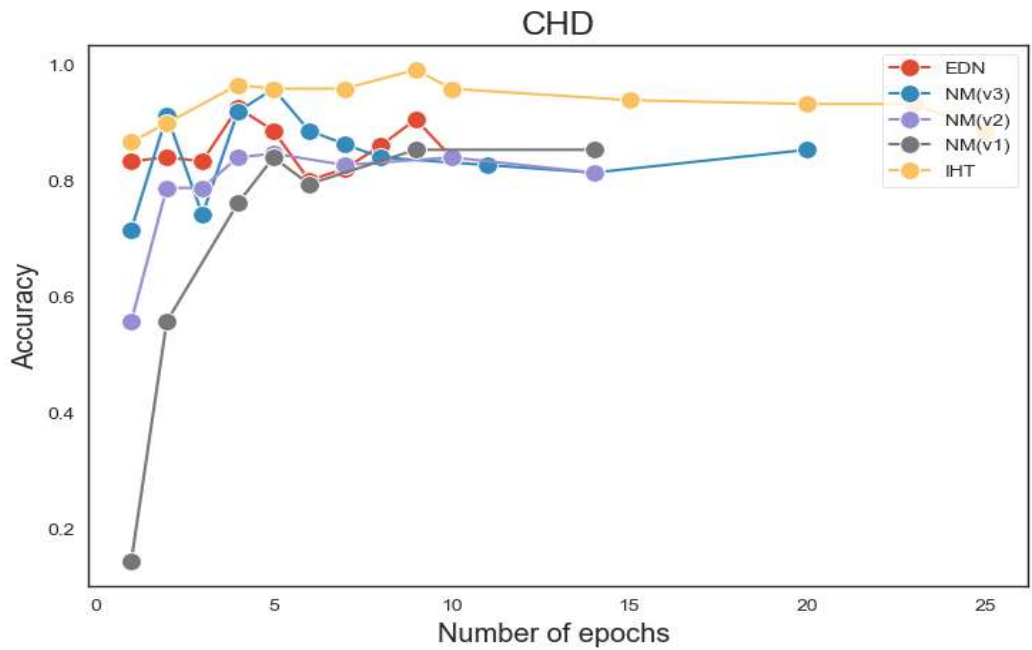

(a)

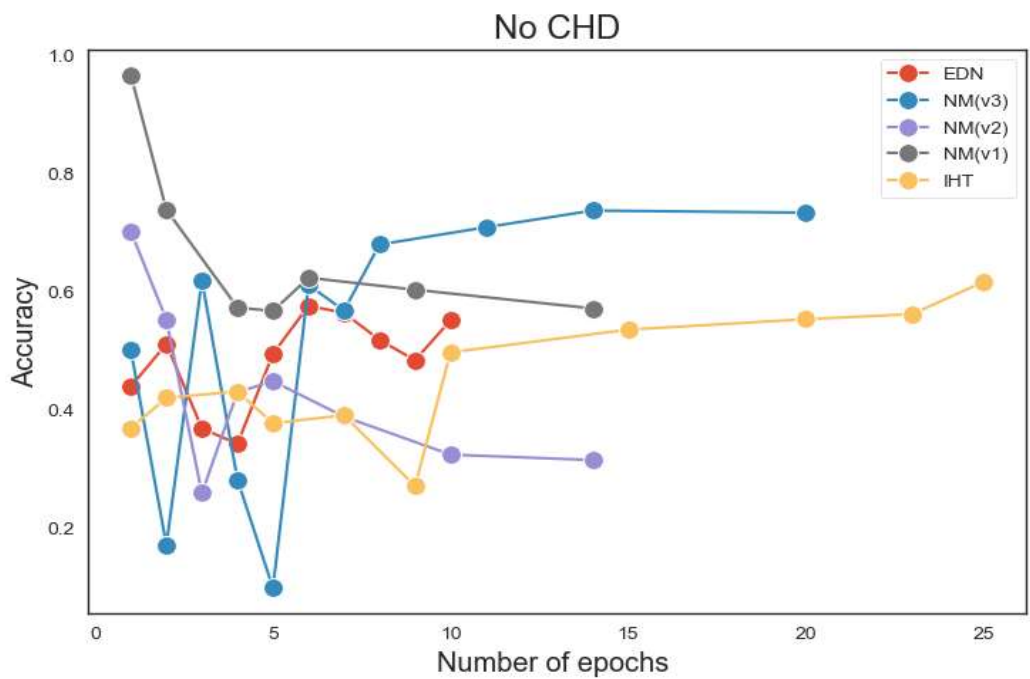

(b)

Fig. 8. Results using the undersampling algorithms from fig. 6 (a)-(e). (a) CHD detection accuracies over epochs using the algorithms. (b) Detection accuracies of no-CHD test data. Among all the competitive undersampling strategies that we compare our results with, near-miss (version 3) works better in improving both class-specific accuracies.

## 4.4. Validation on stroke data

It might appear that sequentially arranged, multiple convolutional layers in our proposed model offer resistance to data imbalance only for the CHD data provided by NHANES. To check whether our model is resilient to other imbalanced datasets containing 1D measurement variables, we apply our network on a similar dataset on Stroke, which is also compiled by NHANES. The Stroke dataset contains 37177 subjects and 36 mixed-type measurement variables. Out of 37177 subjects, there are 1269 subjects who reported that they had strokes. After applying LASSO, we found 34 variables which are important for further processing.

Table VI: Comparison of machine learning models for stroke prediction.

|  | CHD acc (%) | No-CHD acc (%) | Test Accuracy (%) |
|---|---|---|---|
| Logistic Regression | 74 | 76.8 | 76.79 |
| SVM | 75 | 74.75 | 74.74 |
| Random Forest | 74 | 74.45 | 74.44 |
| AdaBoost | 40 | 90.1 | 90 |
| Vanilla-CNN | 71 | 77.04 | 77.01 |
| Our model | 74 | 79.86 | 79.85 |

All the models are trained with data from 1169 patients having CHD and 4300 without CHD. Except our model, we do not apply LASSO prior to training the models. While random forest and SVM yield smaller differences in accuracies between CHD and No-CHD, they fall short in overall test accuracy (74.44% and 74.74% respectively). Our model correctly labels almost 80% of the test cases.

## 4.5. Notes on the resilience to data imbalance

As stated in the introduction, we are primarily interested in problems that follow certain constraints: (1) The data is severely imbalanced due to the nature of its origin or the restriction on its acquisition; (2) Data augmentation techniques, especially that involve under- and over-sampling algorithms and statistical distributions are infeasible; (3) There are significant risks of misclassification. For simplicity, we consider binary classification of data containing mixed-type measurements as variables and investigate the surprising resilience to data imbalance, which is offered by convolution layers. For the assessment of 2D and multidimensional data, we need more refined approaches to address the resilience to data imbalance. It is because in 2D cases, for example image data, there exists spatial correlation among pixels that need to be taken into account, whereas we are considering mixed-type 1D variables that may or may not be correlated at all.

The small number of samples in the minority class and the infeasibility of data augmentation prohibit us from designing deep networks. Therefore, we pay our attention to shallow networks in this context. The results of various sequential convolutional networks are enumerated in Table VII.

Table VII: Experiments with different shallow layers on CHD dataset. I = input, O = output. C2 = a convolution layer with 2 filters, C4 = a convolution layer with four filters, C8 = a convolution layer with 8 filters. 64, 128, 512 = dense layers.

|  | No-CHD acc (%) | CHD acc (%) | Overall acc (%) | Acc difference (%) | No of parameters |
|---|---|---|---|---|---|
| MLP - I (I-512-128-O) | 84.74 | 57.6 | 84.63 | 27.14 | 88706 |
| MLP - II (I-64-128-256-O) | 81.63 | 75.9 | 81.60 | 5.73 | 45442 |
| Conv-1 (I-128-**C4-C8**-O) | 78 | 75.96 | 77.97 | 2.04 | 6306 |
| Conv-II (I-128-**C2-C4**-C8-O) | 79.4 | 77 | 79.38 | 2.4 | 3104 |
| Conv-III (I-128-C4-O) | 77.42 | 75.35 | 77.4 | .07 | 6426 |
| Conv-IV (I-64-C2-128-C4-O) | 81.53 | 76.9 | 81.43 | 4.63 | 11,678 |
| Our model (I-64-**C2-C4**-512-O) | 83.17 | 77.88 | 82.3 | 5.29 | 32066 |

Table VII suggests that, if properly trained, MLP indeed shows improvement in accuracy of the majority class, which unfortunately affects the accuracy of the minority class. The difference between classwise accuracies is 27.14% for MLP-I. MLP-II with an extra deep layer compared to MLP-I seems to have a decline in the accuracy difference (5.73%). However, this is achieved after a careful training and there is a significant chance of overfitting as the number of trainable parameters of MLP-II is 45,442 and this is approximately 9 times of the amount of input data.

It can be noticed that the convolutional layers provide surprising resilience to class imbalance in terms of the difference between classwise accuracies. Conv-I, II and III yield 2.04%, 2.4%, and 0.07% accuracy differences. However, this comes at a cost of achieving lower overall accuracy scores. By restricting ourselves to sequential design for simplicity, we start investigating two possible architectures Conv-IV and the last one in Table VII. After rigorous training, it is observed that sequential placements of C2-C4 outputs better accuracy.

Note that the total number of parameters of our architecture is 32066, which is significantly higher than Conv-I, II and III, but moderately lower than MLP-I and II. Such large number of parameters is due to the presence of the dense layers (64 and 512). While C2-C4 attempts to minimize the difference between classwise accuracies, dense layers try to improve the overall test accuracy.

Another point worth to mention is the stability of accuracy achieved by individual model. While training MLP-I and II, it is observed that the training and test accuracies have the tendency to monotonically increase over epochs when we gradually decrease the weight ratio to 13:40 (see Table III for weight ratio) according to the previously-mentioned training schedule. This is degenerative because after each epoch, the accuracies (train and test) tend to be higher than the ones at previous epoch (destabilization) while the accuracy of the minority class starts plummeting. This degeneration is strikingly diminished after the inclusion of multiple convolution layers. In short, the accuracies that our proposed model yields so far are stable for a fairly large number of epochs.

## 5. Conclusion, Limitations and Future Research

In this paper, we propose a multi-stage model to predict CHD using a highly imbalanced clinical data containing both qualitative and quantitative attributes. For such clinical data, imbalance is an imminent challenge that exists due to the limited availability of data. Such data imbalance adversely affects the performance of any state-of-the-art clinical classification model. As a remedy to the imbalance problem, one cannot efficiently apply conventional techniques, such as data augmentation strategy due to biological implausibility of replication of several attributes in the clinical data. By way of extensive experimentation and validation, we establish that a special-purpose, shallow convolutional neural network exhibits a considerable degree of resilience towards data imbalance, thereby producing classification accuracy superior to the existing machine learning models (Table V). Our model is simple in concept, modular in design, and offers moderate resilience to data imbalance.

The proposed model initiates with the application of LASSO regression in order to identify the contribution of significant variables or attributes in the data variation. Using multiple instances of randomly subsampled datasets, LASSO is performed repeatedly to check the consistency of the variable contribution, which is a crucial step in our algorithm to control the true-negatives of variable selection. Finally, a majority voting algorithm is applied to extract the significant variables of interest, a step that achieves dimensionality reduction by excising unimportant variables. We do not follow the conventional dimension reduction techniques, such as Local Linear Embedding (LLE) and Principal Component Analysis (PCA) because these methods generally provide dimensions that are linear or nonlinear combination of the data variables, leading to a lack of interpretability of the derived dimensions. For example, a linear combination of BMI and alkaline phosphatase (ALP) is difficult to interpret. Rather, we explicitly use t-SNE for the visualization of under and over-sampled data generated by applying state-of-the art algorithms. As we utilize LASSO, a potential research avenue would be to test if LASSO reflects the true importance of a variable through its shrinkage, and if not, this would call for the construction of an appropriate optimization function.

Once we obtain the significant predictor variables with LASSO and feature voting, we feed them to our 1-D feedforward convolutional neural network. We substantiate that shallow convolutional layers provide adaptability in data imbalance in terms of our results in Table V, where, in contrast to Logistic Regression and Adaboost, our model provides a balanced class wise classification accuracy. In a cohort of 37,079 individuals with high imbalance between presence of CHD and Non-CHD, we show that it is possible to predict the CHD cases with 77.3% and Non-CHD cases with 81.8% accuracy, indicating that the prediction of the CHD class, which is deficit in the number of reported patient samples, does not suffer significantly from the data imbalance. The preprocessing stage, consisting of repeated LASSO and majority voting, is pivotal in filtering out highly correlated variables, setting flags only for the uncorrelated ones to be fed to our CNN model. This is clearly observed from Figure 3. Each LASSO stage maintains 1:1 ratio of CHD: Non-CHD data to avoid the adverse effect of data imbalance on the final shrinkage parameters $\gamma$. However, the 1:1 ratio does not encapsulate enough variation in Non-CHD classification as this class contains large number of reported cases in our dataset. We repeat LASSO with randomly sampled subsets and eventually apply majority voting to assess the importance of a variable. Once the dominant predictor variables are identified, we discard their LASSO values in this paper. In future work, instead of disregarding the LASSO-majority voting generated weight values, one can integrate them to the subsequent CNN model as priors, and test if that leads to enhancement in the performance of the model.

A potential problem might arise while using LASSO due to the linear nature of the estimator. LASSO is a penalized regression technique, where sparsity of variables is enforced by a non-convex penalty. Competitive methods, such as cross-correlation based variable selection, also supports the linear map, however, with a little difference. LASSO exploits partial correlation between the factors, which caters to the relevant prediction of output responses, whereas cross-correlation computes, in a sense, the marginal correlation between each pair of factors, which might not be monotonic and linear all the time. Nonetheless, the assumption of linear relationship between the input factors and output labels may have some consequential limitations and the number of factors might be significantly greater than what is present in the current data. A further refined approach, in this case, would be a two-step, non-linear reduction of dimension, where, we can use techniques, such as sure independence screening (SIS), conditional SIS or graphical LASSO to approximate the partial correlation/covariance among the factors. A suitable threshold would give a reduced dimension to apply LASSO afterwards for further reduction in the dimension.

A possible future direction of this work is to consider nutrition and dietary data recorded by NHANES as additional predictor variables for CHD prediction. Dietary factors play an important role in CHD occurrence (Masironi, 1970, Bhupathiraju 2011) and the prediction accuracy of CHD by including additional dietary variables could be explored. For example, until very recently, several prospective studies concluded that total dietary fat was not significantly associated with CHD mortality (Skeaff 2009, Howard 2006). However, according to American Heart Association (AHA), it is the quality of fat which determines CHD risk (Lichtenstein 2006, USDA 2010). Individual experiments performed with NHANES dietary data have discussed the association of cholesterol, LDL, HDL, amino acids and dietary supplements with CHD (references). However, individual consumption of nutrients takes place collectively in the form of meals consisting of combination of nutrients (Hu 2002, Sacks 1995). This may lead to multi-collinearity among factors and thus a more complex dietary pattern analysis, controlling for multicollinearity of CHD associated significant nutrients could lead to a more comprehensive approach to CHD prevention. Additionally, some of the clinical predictor variables included in the classification model of CHD prediction may themselves be impacted by certain dietary habits of patients. Thus, inclusion of dietary data of patients along with clinical predictor variables, in prediction of CHD, can also lead to potential endogeneity issues. However, with appropriate treatment of endogeneity, dietary data inclusion is expected to provide further insights and improved accuracy of CHD diagnosis.

Finally, the preferred selection between data augmentation and data subsampling is much debated and demands attention in this section. Our argument in favor of subsampling is as follows: as observed from the t-SNE figures in the result section, the CHD and no-CHD classes are densely interspersed. Moreover, the class-specific clusters are highly non-convex and extremely hard to separate using naïve nonlinear classifiers. Synthetic data samples using strategies, such as a random sampling on the line connecting an arbitrary pair of data samples (used in SMOTE, ADASYN) might receive the wrong label. It is because of the fact that the newly sampled data sample has the likelihood to be labeled as "0" (for training) if the pair of data samples belongs to class "0". However, the data sample may be biologically implausible or, in case of potential plausibility, may actually be a sample from class "1" as both the classes are densely mixed. Especially, when the data is significantly imbalanced, such as in the case of our data, the number of synthesized data samples of the minority class is large. A countable fraction of such newly synthesized, incorrectly labeled data imposes a large bias on the trained network and increases the probability of misclassification. Therefore, we prefer to adopt the sub-sampling strategy, where the authenticity of data is preserved, barring the measurement and acquisition noise. It is an interesting avenue to explore if the extension of shallow CNN models, in terms of architecture and data sub-sampling, to implementation of neural net-based learning on similar clinical datasets, improves the prediction accuracy of the classification process. As explained earlier, our model can also be used as a transfer learning model and the last two dense layers can be retrained for new data. Thus, a significant future research direction would be to implement CNN for predictions from similar clinical datasets where such imbalanced number of positive and negative classifications exist.

**Acknowledgements**

The authors acknowledge the high-performance computational support from The Center for Advanced Computing (CAC) at Queen's University, Canada and the Center for Research Computing at University of Pittsburgh, USA. This research is not funded by any external research grant.